\documentclass[doublespacing]{elsart}

\usepackage{natbib}
\usepackage{graphicx}
\usepackage{amsmath,amssymb}
\usepackage{lscape} 
\usepackage{epstopdf}
\usepackage{longtable}

\begin{document}

\newcommand{\aap}{A\&AS}           
\newcommand{\aj}{AJ}           
\newcommand{\nat}{Nature}
\newcommand{\apj}{ApJ}
\newcommand{\apjl}{ApJL}           
\newcommand{\icarus}{Icarus}
\newcommand{\apjs}{ApJS}
\newcommand{\pasp}{PASP}
\newcommand{\mnras}{MNRAS}

\begin{frontmatter}

\title{A Derivation of the Luminosity Function of the Kuiper Belt from a Broken Power-Law Size Distribution}

\author[label1,label2]{Wesley C. Fraser}
\ead{wesley.fraser@nrc.ca}
\author[label2,label1]{JJ Kavelaars}

\address[label1]{Dept. of Physics and Astronomy, University of Victoria, Victoria, BC V8W 3P6, Canada}
\address[label2]{Herzberg Institute of Astrophysics, National Research Council of Canada, Victoria, BC V9E 2E7, Canada}

Submitted to {\it Icarus}\\
Submitted May 18, 2008\\

Manuscript Pages: 25\\
Tables: 2\\
Figs: 2

\newpage

\begin{abstract}
We have derived a model of the Kuiper belt luminosity function exhibited by a broken power-law size distribution. This model allows direct comparison of the observed luminosity function to the underlying size distribution. We discuss the importance of the radial distribution model in determining the break diameter. We determine a best-fit break-diameter of the Kuiper belt size-distribution of $30<D_b<90$ km via a maximum-likelihood fit of our model to the observed luminosity function. We also confirm that the observed luminosity function for $m(R)\sim21-28$ is consistent with a broken power-law size distribution, and exhibits a break at $m(R)=26.0^{+0.7}_{-1.8}$.
\end{abstract}


\newpage

\end{frontmatter}

{\bf Proposed Running Head:} Kuiper Belt Luminosity Function.\\

{\bf Editorial Correspondence} \\
Wesley C. Fraser\\
Dept. of Physics and Astronomy\\
University of Victoria\\
Victoria, BC, Canada\\
V8W 3P6

\newpage

\section{Introduction \label{sec:Intro}}
The size distribution (SD) of a belt of planetesimals can provide information on the physical and dynamical properties of those objects. Collisions between belt members causes the SD to evolve over time. Low velocity collisions can cause objects to accrete into larger bodies, while high velocity collisions smash bodies into smaller pieces. A generic outcome of collisional evolution simulations of a belt of planetesimals is a power-law SD with a steep slope for large objects, breaking to a shallower slope at some object diameter. Waves or other shapes can be induced in a size distribution, but the general behavior is that of a broken power-law \citep{Obrien2003}. The location of the transition between the steep and shallow SD slopes (break), as well as the small and large object slopes provides information on the strengths of the colliding bodies, and constrains time-scales of accretion and total mass in the belt region \citep{Safronov1969,Dohnanyi1969,Kenyon2001,Kenyon2002,Kenyon2004,Bottke2005} 

The SD of the Kuiper belt has been inferred from a measurement of the belt's luminosity function (LF) \citep[see for example][(F08) and references therein]{Fraser2008}. F08 show that for the largest objects in the belt (diameter $D\gtrsim 100$km), the SD is well represented by a power-law $N(D)\propto D^{-q}$ with a slope $q\sim4.3$. 

Kuiper belt surveys designed to measure the Kuiper belt object (KBO) faint-end LF ($m(R)\gtrsim 26$), have found evidence to suggest that the LF deviates from a power-law \citep{Bernstein2004}. \citet{Bernstein2004} found that a distribution with a steep slope for bright objects, that ``rolled over" to a shallow slope for faint objects was necessary to describe the substantial dirth of objects with $m(R)\gtrsim 27$. They concluded that the underlying KBO SD transitions (``breaks'') to shallower slopes as predicted from collisional evolution models. Using archival Subaru Suprimecam images, \citet{Fuentes2008} provided a substantial increase in the sample of faint ($m(R)\sim26$) KBOs and concluded that the luminosity function breaks from a steep to shallow slope at $m(R)\sim24.3$.

\citet{Bernstein2004} and \citet{Fuentes2008} use analytic functions to represent the shape of the observed LF. These functional forms were not derived from physical or theoretical properties, making infering the underlying KBO SD based on the reported LF difficult. We present here a new functional form of the LF that results from the observation of a belt of planetesimals exhibiting a broken SD. This function provides a direct interpretation of the observed LF in terms of  the size and radial distributions.

In Section~\ref{sec:derivation}, we derive the new functional form. We fit the function to the available  observations of the Kuiper belt in Section~\ref{sec:fit}, describe what future efforts are necessary to constrain the SD in Section~\ref{sec:errors}, briefly discuss implications of the Kuiper belt SD in Section~\ref{sec:discussion}, and make concluding remarks in Section~\ref{sec:conclusions}..

\section{Broken Power-Law Derivation \label{sec:derivation}}
To determine the shape of the LF exhibited from a belt of planetesimals with a given SD, we first consider a number of objects in a sample represented by

\begin{equation}
N\simeq A\int\int R(r) S\left( D\right) dD dr
\label{eq:Nbasic}
\end{equation} 

\noindent
where $r$ is heliocentric distance, $D$ is object diameter, and $R(r)$ and $S\left(D\right)$ are the radial and SDs of those objects (number of objects with distance $r$ to $r+dr$ or diameter $D$ to $D+dD$). $A$ is a normalization constant to present the equation in number of objects per square degree. Here we have implicitly assumed that the radial and size distributions are separable.

The magnitude of brightness of an object  at a heliocentric distance $r$ in AU, with diameter $D$ in km is well approximated as 

\begin{equation}
m=K+2.5\log_{10}\left(r^4 D^{-2}\right)
\label{eq:magnitude}
\end{equation}

\noindent
 where, for KBOs we have applied the simplification, $\Delta \sim r$. Setting $K=18.4$, corresponds to KBOs with 6\% albedos when observed at opposition \citep{Gladman2001}; implicitly assuming that all objects have similar albedos. The largest object observable at a given distance, $r$ in a survey that is sensitive to source magnitudes between $m_{min}$ and $m_{max}$ has diameter $D_{max}=r^2 10^{(m_{min}-K)/5}$. Similarly, the smallest observable object has diameter $D_{min}=r^2 10^{(m_{max}-K)/5}$. 

The cumulative LF of a belt of planetesimals as observed by a particular survey, can be derived from Eq.~\ref{eq:Nbasic} by using the integration limits $D_{min}$ and $D_{max}$. If we assume $S(D) \propto D^{-q}$, ie. we assume a SD is a power-law with `slope' $q$, Eq.~\ref{eq:Nbasic} becomes

\begin{equation}
N\left(<m_{max}\right) = A \int \frac{r^2 R(r)}{1-q} dr \left[ 10^{\frac{K-m_{min}}{5}(1-q)} - 10^{\frac{K-m_{max}}{5}(1-q)}\right]
\label{eq:Bigpwl_qc}
\end{equation}

\noindent
where $q\neq 1$. Setting $\alpha =\frac{q-1}{5}$, $a_1 = A \int \frac{r^2 R(r)}{q-1} dr$, and restricting to the case, $10^\frac{K-m_{min}}{1-q}  \ll  10^\frac{K-m_{max}}{1-q}$, then the cumulative LF can be represented by

\begin{equation}
N\left(m_{max}\right) = a_1\left(m_{max}\right) 10^{\alpha m_{max}}.
\label{eq:full_1}
\end{equation}

We consider the case where the SD breaks to a different slope at a particular break size $D_b$. That is, 

\begin{equation}
S(D) \propto
\begin{cases}
D^{-q_1} & \mbox{for $D>D_b$} \\
D^{-q_2} & \mbox{for $D<D_b$} .
\end{cases}
\label{eq:BrokenSizeDist}
\end{equation}

\noindent
If $r_1$ and $r_2$ define the inner and outer edges of the belt, then a survey where $D_{min}=r_1^2 10^{(m_{max}-K)/5} > D_b$, the observed LF is given by Eq.~\ref{eq:full_1} where the slope $\alpha=\alpha_1=\frac{q_1-1}{5}$. If however, the survey is sensitive to objects smaller than $D_b$ out to some distance $r_b$, then Eq.~\ref{eq:Nbasic} becomes

\begin{equation}
N=A\int_{r_{1}}^{r_{b}} R(r) \left[ \beta \int_{D_{min}}^{D_{b}} D^{-q_{2}} dD + \int_{D_{b}}^{D_{max}} D^{-q_{1}} dD \right] dr  +A\int_{r_{b}}^{r_{2}} R(r) \int_{D_{min}}^{D_{max}} D^{-q_{1}} dD dr
\label{eq:N_mid}
\end{equation}

\noindent
where $\beta$ is a scaling factor to ensure that $S(D)$ is piecewise continuous at $D_b$.

%
%

Using Eq.~\ref{eq:magnitude},  the maximum distance $r_b$ a survey is sensitive to the break size $D_b$, is given by $r_b = D_b^{\frac{1}{2}} 10^{(m_{max}-K)/10}$. 

The constant $\beta$ can be solved for by requiring that the number of objects slightly larger than $D_b$ be equal to the number of objects slightly smaller than $D_b$. That is

\begin{equation}
\beta \int_{D_b-\delta}^{D_b} D^{-q_{2}} dD = \int_{D_b}^{D_b+\delta} D^{-q_{1}} dD,
\label{eq:betarelation}
\end{equation}

\noindent
which gives

\begin{equation}
\beta = D_{b}^{q_2 - q_1}.
\label{eq:beta}
\end{equation}

\noindent
Restricting to the case where $10^{\frac{K-m_{min}}{5}(1-q_1)} \ll 10^{\frac{K-m_{max}}{5}(1-q_1)}$ and $10^{\frac{K-m_{min}}{5}(1-q_1)} \ll 10^{\frac{K-m_{max}}{5}(1-q_2)}$, then the cumulative LF is given by

\begin{equation}
N\left(<m_{max}\right) \simeq  b_1 10^{\frac{q_2-1}{5} m_{max}} + b_2 10^{\frac{q_1-1}{5} m_{max}}  + b_3\left(m_{max}\right)
\label{eq:Fraser}
\end{equation}

\noindent
where 

\[b_1= A \int_{r_1}^{D_{b}^{1/2} 10^{(m_{max}-K)/10}} R(r) D_{b}^{q_2-q_1} \frac{r^{2(1-q_2)}}{q_2-1} dr 10^{\frac{-K}{5}(q_2-1)},\] 
\[b_2= A \int_{r_b}^{r_2} R(r) \frac{r^{2(1-q_1)}}{q_1-1} dr 10^{\frac{-K}{5}(q_1-1)}\]
\noindent
 and 
\[b_3 = A \int_{r_1}^{D_{b}^{1/2} 10^{(m_{max}-K)/10}} R(r) dr D_{b}^{1-q_1} \left[\frac{q_2-q_1}{\left(q_1-1 \right)\left(q_2 -1\right)}\right],\]
 where $q_1 \neq 1$ and $q_2 \neq 1$, and $A$ is chosen such that $N$ is the number of objects per square-degree of the sky.

If a survey is deep enough to see objects with diameters $D \leq D_{b}$ for all $r$, then the cumulative LF then becomes

\begin{equation}
N(<m_{max}) \simeq c_110^{\frac{q_2-1}{5}  m_{max}} +c_2
\label{eq:Fraserfaint}
\end{equation}

\noindent
where 
\[c_1 = A \int_{r_1}^{r_2} R(r) \frac{r^{2\left(1-q_2\right)}}{q_2-1} D_{b}^{q_2-q_1} dr 10^{\frac{-K\left(q_2-1\right)}{5}}\]  \noindent and \[c_2 = A \int_{r_1}^{r_2} R(r) dr D_{b}^{1-q_1} \frac{q_2-q_1}{\left(1-q_2\right)\left(1-q_1\right)}.\]

The cumulative LF of a belt of planetesimals with a two sloped power-law SD is a power-law with slope $\alpha_1$ for magnitudes brightward of $m_1$. The roll-over in the LF begins at the brightest magnitude at which objects smaller than $D_b$ can be detected at the inner edge of the belt, $r_1$. Thus, $m_1=K+2.5log\left(r_1^4 D_b^{-2}\right)$. The LF becomes a power-law again at the magnitude $m_2$ faint-ward of which, objects larger than the break-size are no longer observed at the outer edge of the belt, $r_2$. Thus, $m_2=K+2.5log\left(r_2^4 D_b^{-2}\right)$. For magnitudes $m_1\leq m\leq m_2$, the LF is the sum of nearby objects with $D<D_b$, and distant objects with $D> D_b$. Hence both the radial and size distributions become important as the shape of the LF is determined by the coupling of the two. The functional form of this LF is

\begin{equation}
N\left(<m\right) = A
\begin{cases}
 a_1 10^{\alpha_1 m} & \mbox{if $m < m_1$}\\
 b_110^{\alpha_2 m} + b_2 10^{\alpha_1 m} + b_3\left(m\right) & \mbox{ if $m_1 \leq m \leq m_2$}\\
 c_1  +c_2 10^{\alpha_2 m} & \mbox{ if $m > m_2$}.
 \end{cases}
 \label{eq:Lum_func_simple}
 \end{equation}
 
\noindent
Comparing to Eq.~\ref{eq:Fraser}, we see that the SD slopes $q_1$ and $q_2$ are related to the LF slopes $\alpha_1$ and $\alpha_2$ by $q_1=5\alpha_1+1$ and $q_2=5\alpha_2+1$.

The power-law coefficients, $a_i, b_i, \mbox{ and } c_i$, of this function depend on the radial distribution. For the case where a break is observed, knowledge of the true radial distribution is important as the shape of the LF is determined from the coupling of the radial and size distributions. For the case where the break is not seen, the radial distribution is not important as it only affects the LF normalization. The size distribution slopes $q_1$ and $q_2$ can be inferred from the LF slopes $\alpha_1$ and $\alpha_2$ if the LF has been observed sufficiently brightward and faint-ward of the  roll-over.


 \section{Model fit to the Observations \label{sec:fit}}
We present here a fit of our model (Eq.~\ref{eq:Lum_func_simple}) to the observed KBO LF. 

Eq.~\ref{eq:Nbasic} can be evaluated analytically if $R(r) \propto r^{-c}$ where $c$ is the radial distribution `slope'. While this is likely not the true KBO radial distribution, \citet{Trujillo2001} and \citet{Kavelaars2008} have shown that the radial distribution is sharply decreasing beyond $r\sim 40$ AU. A visual examination of Figure. 2 from \citet{Trujillo2001} reveals that $R(r)=\left(\frac{r}{63}\right)^{-10}$ is a good representation of the sharp density fall-off. Such a representation is sufficient to show the importance of accurately modeling the radial distribution when determining the size distribution. The full functional form using this radial distribution is presented in the appendix.


\citet{Fraser2008} present a fit of the LF for data brightward of $m(R)=26$. They find that the LF of the Kuiper belt is well described by a single power-law with slope $\alpha=0.65 \pm 0.05$. To ensure reliability of the fit, they only considered surveys with measured detection efficiencies as a function of magnitude for each field in the survey. In fitting Eq.~\ref{eq:Lum_func_simple} to the observed LF, we adopted the same practices as \citet{Fraser2008} listed below:

\begin{enumerate}
\item We considered the F08 sample presented by \citet{Fraser2008}. In addition, we included the survey presented by \citet{Bernstein2004,Bernstein2005}. Note: we did not included the survey presented by \citet{Fuentes2008} as the majority of analysis for this manuscript was complete before the author was aware of that work. This manuscript focuses on the functional form of the LF, and these data, which may improve the constraint on the function parameters, do not change our conclusion that a full understanding of the size distribution requires a rigorous treatment of the radial distribution as well.
\item We offset all magnitudes to R-band using the average KBO colours presented in \citet{Fraser2008}.
\item We cull from each survey all sources faintward of the 50\% detection efficiency of that survey, and set the detection efficiency to zero faintward of that point.
\item We fit the differential LF, $\Sigma(m)=\frac{dN(<m)}{dm}$ using a maximum likelihood technique.
\item We adopted a form of the likelihood equation as derived by \citet{Loredo2004}, and extended to account for sky density variations, $A_k$, and color variations, $C_k$ \citep[see][for details]{Fraser2008}. We adopted the same prior ranges of the colour offsets and range of normalization parameters as presented in \citet{Fraser2008} ($\pm 0.2$ for the colour parameters, and $\pm 0.4$ for the normalization parameters). 
\end{enumerate}

To evaluate Eq.~\ref{eq:Lum_func_simple}, we require a set of radial distribution parameters, $r_1, r_2, \gamma=(1-c)/10$. We set the Kuiper belt inner edge as $r_1=35$ AU.  For a baseline model, we consider an outer edge $r_2=60$ AU with the radial fall-off parameter, $c=10$ (see previous discussion). This provides a reasonable representation of the steep fall-off in the Kuiper belt radial distribution and typical outer limit to the objects detected in the F08b sample. We also consider two test cases; in model 2 we set $r_2=100$ AU, and in model 3, we set the radial fall-off parameter, $c$, to that of the minimum-mass solar nebula, $c=3/2$ \citep{Hayashi1981}  corresponding to $\gamma=-0.05$. Models 2 and 3 were chosen to evaluate the effects of the radial distribution parameters on the fit results. We speculate that the value of $r_2$ for any reasonable choice, will have no strong effect on the fit parameters as most objects are detected at distances less than $\sim45$ AU.


Using our maximum likelihood technique, we determined the best-fit LF parameters of Eq.~\ref{eq:Lum_func_simple} using the baseline model, as well as models 2 and 3. These are presented in Table~\ref{tab:fit}. The observations were best-fit with $\alpha_1=0.69\pm0.07$, $D_b=37_{-10}^{+50}$ km, $\alpha_2=-0.4_{-3}^{+0.8}$ for our baseline model. 

We found that the best-fit $\alpha_1$ and $D_b$, $D_b=37_{-10}^{+50}$ km and $\alpha_1=0.69\pm0.07$, are not dependent on the choice of radial distribution models. As demonstrated in Eqs.~\ref{eq:Bigpwl_qc} and \ref{eq:Fraser}, the radial distribution has no effect on the apparent bright object slope and the break diameter depends only on the choice $r_1$. Thus, because the same data was always used, and because we set $r_1=35$ AU for all choices of radial distribution, the best-fit $\alpha_1$ and $D_b$ {\it should} be the same for all radial distributions considered.  

We found very different behaviors for the faint-end slope from our two choices of $\gamma$.  The best-fit $\alpha_2$ was found to be $-0.4_{-3}^{+0.8}$, $-0.4_{-3}^{+0.8}$, and $-\infty$ for the baseline model $(r_1,r_2,\gamma)=(35,60,-0.9)$, Model 2 $(r_1,r_2,\gamma)=(35,100,-0.9)$, and Model 3 $(r_1,r_2,\gamma)=(35,100,-0.05)$ respectively. 

With $\gamma=-0.05$ ($c=3/2$), we found that the observations were best-fit with a SD in which there is a complete absence of objects, ($\alpha_2=-\infty$) smaller than break diameter, $D_b\sim 40$ km. When $\gamma=-0.9$ ($c=10$), the best-fit faint end slope was $\alpha=-0.4$. These two behaviors are expected; a sharp roll-over in the LF can be the result of a very sharp roll-over in the SD with a gradual fall-off in the radial distribution, or a less sharp roll-over in the SD and a steep fall-off in the radial distribution. Setting $c=3/2$ does not correctly model the very sharp radial fall-off in the Kuiper belt and thus it is unlikely that there is a complete absence of objects smaller than $\sim D_b$. We discard model 3 as this is an unlikely model of the radial and size distributions. We do note however, that no KBOs smaller than $D\sim30$ km have yet been detected.

We found no difference in the best-fit parameters when $r_2$ was varied (see the baseline model and model 2). This demonstrates the expected lack of influence of this parameter on the fits. 

A series of Monte-Carlo simulations were performed in which a sample of detections representative of the surveys and detections in the F08 sample were generated from a model Kuiper belt. Each simulated data-set was fit using the maximum likelihood technique, and on average, the fitting routine reproduced the input model parameters. The same behavior was found for the fits of the simulated observations, as for the real observations. That is, the best-fit faint-end slope, $\alpha_2$, depended on our choice of $\gamma$, and the best-fit break diameter, $D_b$, depended on our choice of $r_1$. The reported best-fit values were correct however, if the chosen $\gamma$ and $r_1$ were within a factor of $~2$ the correct value. Thus we feel that our likelihood method presented here produces satisfactory results to within the accuracy of the fit to the data currently available.

All minimizations have the same maximum likelihood value to within a few percent. Thus there is no preference, in a statistical sense, to choose one model over another; each is equally sufficient to describe the observations. Thus, we take take the baseline model, with $(r_1,r_2,\gamma) = (35,60,-0.9)$ and $(\log A,\alpha_1,\alpha_2,D_b)=(22.82,0.69,-0.4,37)$ as the best model, as the baseline model radial distribution is the most similar of the three considered, to that observed for the Kuiper belt. 

The differential LFs (derivative of Eq.~\ref{eq:Lum_func_simple} wrt. $m$) using the best-fit parameters for all choices in radial distribution are presented in Fig.~\ref{fig:cumulative}. To produce the histogram, object magnitudes and the areas of each survey have been adjusted using the best-fit $C_k$ and $A_k$ parameters (see F08 for details). These parameter values were determined by maximizing the likelihood while holding the LF parameters to their best-fit values, and are presented in Table~\ref{tab:colours}. The histogram is presented in order to visualize the observations, but was never used in the fitting procedure. As can be seen, a very sharp roll-over is apparent for magnitudes faint-ward of $m(R)\sim 26$. This plot provides visual confirmation that the fit describes the data reasonably well.

The likelihood contours of the best-fit LF for our choice in baseline radial distribution is presented in Fig.~\ref{fig:likelihood}. As can be seen, $\alpha_1$ and $A$ and $\alpha_2$ and $D_b$ are highly correlated. The uncertainties of the best-fit parameters in Table~\ref{tab:fit} are the extrema of the 67\% bayesian credible regions. 

The best-fit large object slope and break diameter, $q_1=4.45\pm0.35$ and $Db=37_{-10}^{+50}$,  are consistent with \citep{Fraser2008} who found that $q=4.25\pm0.75$ for objects larger than $D\sim 50$ km. 

\section{Sources of Uncertainty \label{sec:errors}}
The best-fit LF we present assumes that the lack of faint KBOs observed by \citet{Bernstein2004} and  \citet{Fuentes2008} is caused by a break in the SD, and suggests that the break observed must occur at diameter $27< D_b < 87$. This result however assumes that an R-albedo of 6\% is typical of KBOs. Recent observations suggest that this is true for objects $D\lesssim500$ km, but objects with $D\gtrsim 1000$ km have albedos $\sim60-90$\% \citep{Grundy2005,Stansberry2007}. \citet{Fraser2008} has shown however, that the trend of lower albedos for smaller objects would produce a {\it steeper} LF than otherwise would be observed for constant albedos. Thus, albedo effects are likely not the source of the observed break. In addition, \citet{Fraser2008} found that a variation of albedo with distance has no effect on the inferred LF slopes, and only the LF normalization.

We also assume that the size distribution is the same at all distances. If this is not true, it would have similar affects as assuming constant albedos for all sizes. That is, the slope inferred under the assumption that the size distribution is distance dependent  could be steeper or shallower than the true slope depending on the trend of size with distance. This effect cannot be tested without substantially more observations than currently available. This effect however, is likely negligible, as  KBOs have excited orbits. Thus interactions (collisions) between objects occur over a range of many tens of AU which would cause the size distribution to be similar over the full range of the Kuiper belt.

The LF also assumes that the break in the SD is sharp. That is, there is no size range over which the SD slope transitions from $q_1$ to $q_2$. While this is certainly not true, simulations of accretion and collisional processing in a belt of planetesimals show that the transition occurs over a small range of sizes compared to the range over which the SD exhibits a power-law like behavior \citep{Kenyon2002}. Additionally,  the lack of surveys sensitive to KBOs with $m(R)>27$ from which the roll-over shape is constrained implies a more complicated description is not warranted. More survey data is needed that covers a few square degrees in the range $m(R)\sim 26-28$ to accurately constrain the break magnitude; a large  number ($N \sim 20$) of detected objects with a range of magnitudes beyond the break are necessary to constrain the faint end slope, without which the break diameter determination remains uncertain. 
 
 The SD inferred from the LF is still model dependent, and determining the structure of the SD reliably from the LF requires improved knowledge of the properties of the Kuiper belt (albedo, radial distribution, etc.). A large source of uncertainty stems from the current lack of knowledge of the KBO radial distribution. Each survey considered in this analysis sampled the Kuiper belt at different locations on the sky. Variations in the sky density between surveys, requires treating sky density parameters as nuisance parameters for each individual field \citep{Fraser2008}. This causes the range of parameter values consistent with the observations to be grossly enlarged. An accurate Kuiper belt model which provides the instantaneous radial distribution as a function of latitude and longitude is needed such that density estimates and the coefficients of Eq.~\ref{eq:Lum_func_simple} can be calculated as a function of latitude and longitude. This would remove the need for treating sky densities as nuisance parameters, and would greatly reduce the uncertainty in the inferred SD.

\section{Discussion \label{sec:discussion}}

Our results have demonstrated that the size distribution of the Kuiper belt can be described by three size distribution parameters, $q_1$, $q_2$, and $D_b$, and that, to accurately measure the break-slope $q_2$ requires that the radial distribution be accounted for in a way which removes the ambiguity between source brightness and source size. We have found that $q_1\sim4$, $q_2 \sim -1$, and $D_b\sim40$ km.

As in previous works, we have found that the observed large object slope, $q_1\sim4$ \citep{Trujillo2001b,Gladman2001,Bernstein2004,Petit2006,Fraser2008}, is inconsistent with the $q\sim 3.5$ slope expected from a steady-state collisional cascade \citep{Dohnanyi1969,Obrien2003,Pan2005}. Rather, the slope is consistent with the large object slope produced from accretionary processes \citep{Safronov1969,Kenyon2001,Kenyon2002}.

Various accretion and collisional models exist that can reproduce some of the observed features of the size distribution, ie. the correct small object slope, or break diameter. But to this date, none can simultaneously reproduce $q_1$, $q_2$, and $D_b$. For instance,  \citet{Pan2005} present a collisional model in which the accretion size distribution breaks to one in a state of collisional equilibrium. They demonstrate that a break diameter consistent with our results could occur if collisional disruption occurred for timescales $\sim 1-4$ Gyr. The equilibrium slope they considered $(q_2\sim3)$ however, is inconsistent with the small object slope of our baseline model. The assumption \citet{Pan2005} make, that the size distribution immediately breaks from an accretionary slope to collisional equilibrium is likely incorrect, as there is likely a range of sizes over which disruptions by the more numerous smaller objects have significantly depleted the population, but which have not been replenished with fragments from larger object disruptions. This transition range is seen in more complicated accretion models \citep{Kenyon2001, Bottke2005}.

In early times of the accretion simulations of \citet{Kenyon2001,Kenyon2002}, steep large object slopes, $q_1\sim4$ are produced. For times longer than $\sim 70$ Myr ($\sim 1$ Gyr for weak bodies), the slope rapidly evolves to a shallower slope inconsistent with our findings. This suggests that accretion in the primordial Kuiper belt was a short-lived process which must have been `turned off' by some process - likely the dynamical excitation of the belt by some scattering process, such as gravitational stirring from a migrating or scattering Neptune \citep{Malhotra1993,Thommes2002,Levison2003} or a rogue planet \citep{Gladman2006}. These simulations also produce a break diameter, $D_b\sim 10$ km - only moderately consistent with our results. \citet{Kenyon2004} found that, with gravitational stirring from Neptune, after 4.5 Gyr, `weak' KBOs - that is, bodies with collisional strengths lower than expected from icy planetesimals - exhibited a roll-over at break diameters as large as $D_b\sim 30$ km, consistent with that found here. In addition, there simulations produced small object slopes consistent with our findings, but produced a large object slope which is too shallow compared to the observed slope.

The models studied by \citet{Kenyon2001,Kenyon2002,Kenyon2004} have suggest that KBOs are effectively strengthless, gravitationally bound rubble-piles. These models however, cannot simultaneously produce both the large object and small object slopes; either the evolution is short enough such that the steep accretion slope is maintained, or the evolution is long enough such that a large enough break, and shallow enough small object slope are produced. This is not a surprise however, as it is apparent that the Kuiper belt history is not as quiescent as that considered in these models. The highly excited KBO orbital distribution cannot be produced by in-situ formation. Some large-scale scattering event is needed. Thus, the correct dynamical history must be determined before reliable conclusions can be drawn about the accretion/disruption history, and strengths of KBOs.

\section{Conclusions \label{sec:conclusions}}
We have developed a model of an LF exhibited from a planetesimal belt with a broken power-law SD. This model allows a direct interpretation of the observed LF in terms of the underlying SD assuming a broken power-law shape. We fit this model to the observed Kuiper belt LF and the break first observed by \citet{Bernstein2004}. The best-fit parameters are presented in Table~\ref{tab:fit}. We find that the break in the SD must occur at diameters $27-87$ km ($m(R)\sim25.8$) for reasonable choices of the radial distribution. We have demonstrated that the inference of the SD from the LF requires an accurate model of the radial distribution. Additional detections of KBOs faintward of $m(R)=26$, and a proper treatment of the KBO orbital distribution are required to provide an accurate measure of the break size and faint-end slope.

\section{Acknowledgements}
The authors would like to thank the reviewers for their helpful comments which patched a few holes in the original manuscript not recognized by the authors. A special thanks goes to Doug Johnstone for his very careful read through of this manuscript. This project was funded by the National Science and Engineering
Research Council and the National Research Council of
Canada. This research used the facilities of the Canadian Astronomy
Data Centre operated by the National Research Council
of Canada with the support of the Canadian Space Agency.

\section{Appendix}
As an approximation to the true Kuiper belt radial distribution, we assume that $R(r)\propto r^{-c}$ and $r_1\leq r \leq r_2$.  Making the substitution $\gamma = (1-c)/10$, and collecting like terms, the coefficients of Eq.~\ref{eq:Lum_func_simple} become

\begin{equation}
a_{1} = A \frac{r_{2}^{10(\gamma-\alpha_1)} - r_{1}^{10(\gamma-\alpha)}}{50\alpha(\gamma-\alpha)}10^{-\alpha K},
\end{equation}

\begin{equation}
b_{1} \sim -A\frac{D_{b}^{5\left(\alpha_2-\alpha_1\right)}}{50\alpha_2\left(\gamma-\alpha_2\right)} r_1^{10\left(\gamma-\alpha_2\right)} 10^{-K\alpha_2},
\end{equation}

\begin{equation}
b_{2} \sim A\frac{r_{2}^{10(\gamma-\alpha_1)} }{50\alpha_1(\gamma-\alpha_1)} 10^{-\alpha_1K},
\end{equation}

\begin{equation}
b_{3} (m) \sim A \frac{\left(\alpha_2 - \alpha_1\right)}{\left(\gamma-\alpha_1\right) \left(\gamma-\alpha_2\right)} \frac{D_{B}^{5\left(\gamma-\alpha_1\right)}}{50\gamma}10^{\gamma \left(m-K\right)}
\end{equation}

\begin{equation}
c_{1} = A \frac{r_{2}^{10\gamma}-r_{1}^{10\gamma}}{10\gamma} D_{b}^{-5\alpha_1}\left(\frac{\alpha_2-\alpha_1}{5\alpha_1\alpha_2}\right),
\end{equation}

\noindent
and

\begin{equation}
c_{2} = A\frac{r_{2}^{10(\gamma-\alpha_2)}-r_{1}^{50(\gamma-\alpha_2)}}{50\alpha_2(\gamma-\alpha_2)} D_{b}^{5(\alpha_2-\alpha_1)} 10^{-\alpha_2K}.
\end{equation}

\noindent
where $\gamma \neq \alpha_1$, $\gamma \neq \alpha_2$, $\alpha_1 \neq 0$, and $\alpha_2 \neq 0$, or equivalently $3-2q_1-c \neq 0$, $3 -2q_2-c\neq 0$, $1-q_1 \neq 0$, and $1-q_2\neq 0$ and we have dropped the negligible terms from the coefficients.

\bibliographystyle{icarus}
\bibliography{AstroElsart}

\newpage

\begin{table}[h]
   \caption{Maximum Likelihood Parameters. The bright and faint object slopes $\alpha_1$, $\alpha_2$, the break diameter $D_b$, and the normalization constant $A$ were fit by the maximum likelihood routine. Belt edges $r_1$ and $r_2$ as well as the radial fall-off parameter $\gamma$ are fixed parameters not fit by the maximum likelihood routine. Fit 1 represents our baseline model, as the chosen radial distribution best matches that of the Kuiper belt. Fits 2 and 3 have different values of $r_2$ and $\gamma$ respectively, chosen to determine the sensitivity of different choices in radial distribution models on the measured LF parameters. Note: $\gamma=-0.05$ produces physical implausible results. ie. $\alpha_2=-\infty$. \label{tab:fit}}\begin{tabular}{cccc}
	Parameter & Baseline Model & Model 2 & Model 3 \\ \hline
         $r_1$ & $35.0$ & $35.0$& $35.0$\\
         $r_2$ & $60.0$ & $100.0$&$100.0$\\
         $\gamma$ & $-0.9$ & $-0.9$&$-0.05$\\
	$\log A$ & $ 22.8\pm0.6 $ &$22.8\pm0.6$& $9.3\pm0.6$\\ 
	$\alpha_1$  & $0.69\pm0.07$ &$0.69\pm0.07$ & $0.69\pm0.07$\\
	$\alpha_2$  & $-0.4^{+0.8}_{-3}$ &$-0.4^{+0.8}_{-3}$ & $-\infty$\\
	$D_b$ (km)  & $37^{+50}_{-10}$  &$37^{+50}_{-10}$& $37^{+50}_{-10}$  \\
\end{tabular}
\end{table}

\newpage
\begin{longtable}{cccc}
	Field & $\log A_k$ & $C_k$ & Source\\ \hline
	UN & 22.6 & 0.0&  \citet{Fraser2008}\\
	MEGA & 23.0 & -0.15&  \citet{Fraser2008} \\	
	CTIO01 & 23.2 & 0.1&   \citet{Fraser2008} \\	
	CTIO02 & 23.1 & 0.05&   \citet{Fraser2008} \\	
	SSU & 23.0 & 0.0& \citet{Petit2006} \\	
	SSN & 23.1 & 0.03& \citet{Petit2006} \\	
	G01 & 22.9 & -0.01& \citet{Gladman2001} \\	
	AF & 23.0 & 0.0&  \citet{Allen2002} \\	
	AKL & 23.2 & 0.0& \citet{Allen2002} \\	
	TE1G & 22.8 & 0.0& \citet{Trujillo2001b} (see \citet{Fraser2008}) \\	
	TE2G & 22.8 & 0.0& \citet{Trujillo2001b} (see \citet{Fraser2008}) \\	
	TE3G & 22.7 & 0.0& \citet{Trujillo2001b} (see \citet{Fraser2008}) \\	
	TE5M & 22.8 & 0.2& \citet{Trujillo2001b} (see \citet{Fraser2008}) \\	
	Bern & 22.7 & -0.2& \citet{Bernstein2004} \\	
	KeckV & 22.6 & 0.2& \citet{Chiang1999} \\	
	JLT98 & 22.5 & -0.2& \citet{Jewitt1998}  \\	
	G98UH & 22.4 & -0.1& \citet{Gladman1998} \\	
	G98Palomar & 23.2 & -0.03& \citet{Gladman1998} \\	
	TE1M & 22.8 & 0.13& \citet{Trujillo2001b} (see \citet{Fraser2008}) \\	
	TE3M & 22.9 & 0.19& \citet{Trujillo2001b} (see \citet{Fraser2008}) \\	
	TE4G & 22.8 & 0.0& \citet{Trujillo2001b} (see \citet{Fraser2008}) \\	
	TE4M & 22.7 & 0.05& \citet{Trujillo2001b} (see \citet{Fraser2008}) \\	
	AB & 22.9 & -0.11& \citet{Allen2002} \\	
	AD & 22.8 & 0.0& \citet{Allen2002} \\	
	AE & 22.9 & 0.0& \citet{Allen2002} \\	
	AG & 22.5 & 0.0& \citet{Allen2002} \\	
	AJ & 22.8 & 0.08 & \citet{Allen2002}\\
   \caption{Best-fit normalization parameters $A_k$ and colour off-set parameters $C_k$ for each field as defined in \citet{Fraser2008}. These are treated as nuisance parameters and are marginalized when determining the best-fit LF (see Section~\ref{sec:fit}). For this table, and Fig.~\ref{fig:cumulative}, these best-fit parameter values are determined from the maximum likelihood routine when the LF parameters are set to their best-fit values (Fit 1 from Table. \ref{tab:fit}). \label{tab:colours}}
\end{longtable}

\section*{Fig. Captions}
\begin{figure}[h] 
   \centering
   \includegraphics[width=6in]{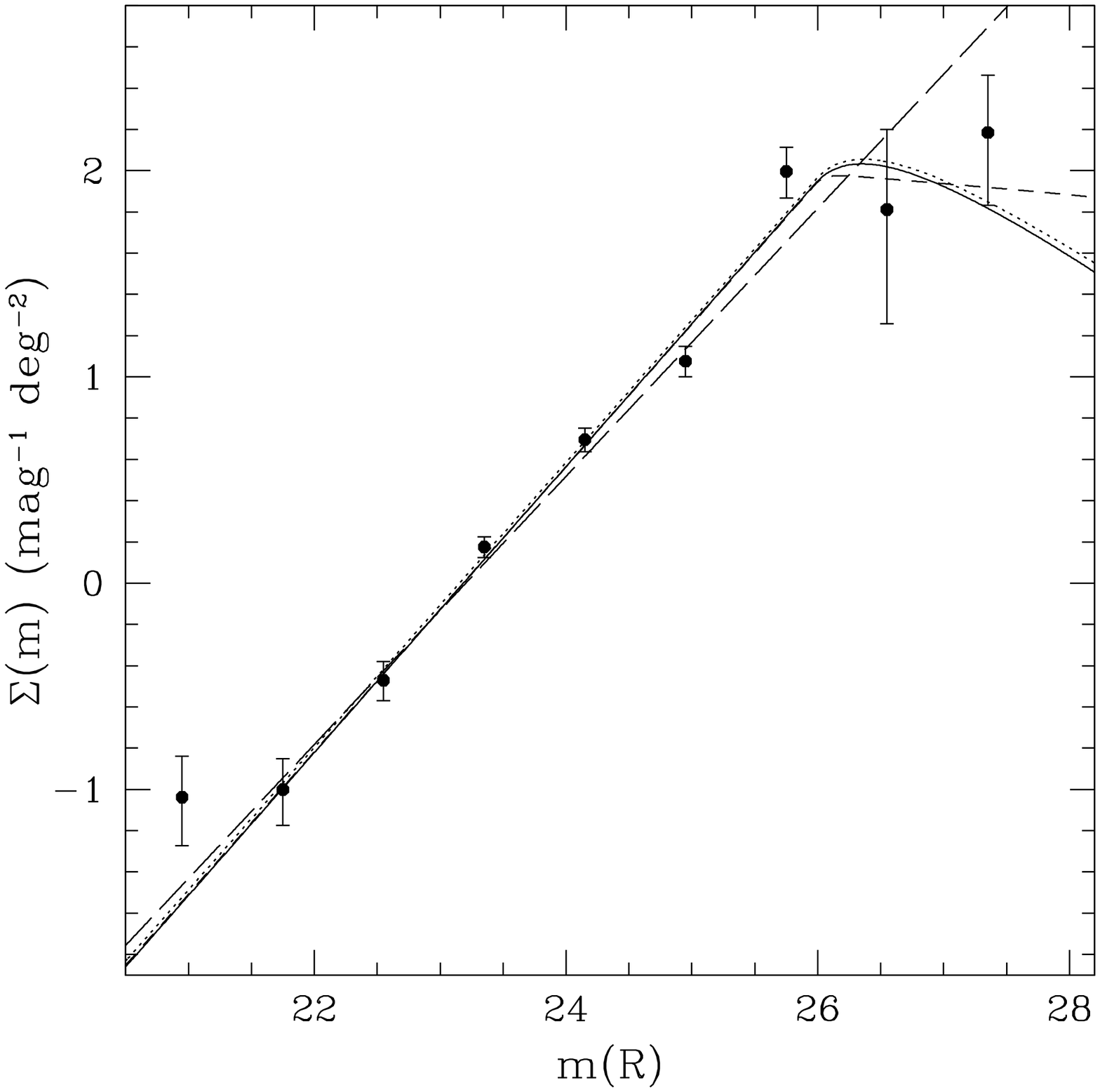} 
   \caption{ \label{fig:cumulative}}   
\end{figure}

Fig.~\ref{fig:cumulative}~Differential surface density histogram of data from the F08 sample and \citet{Bernstein2004} using 0.8 mag bin-widths. Objects magnitudes have been shifted using the best fit colour offsets $C_k$, and observed sky densities have been adjusted using the normalization values $A_k$ taken from Table~\ref{tab:colours}. Dashed straight line: best fit power-law with $\alpha = 0.65$ and $m_{R}=23.42$ taken from \citet{Fraser2008}. Solid curve: Model 1 (baseline model), dotted curve: Model 2, and  dashed curve: Model 3. The baseline model has the radial distribution which best represents that of the Kuiper belt. Models 2 and 3 test the sensitivity of the fit to different radial distributions. Parameters listed in Table~\ref{tab:fit}. As can be seen, a strong break is seen in the LF, as a deviation from the power-law behavior faint-ward of $m(R)\sim25.8$. The shape of the binned LF is highly sensitive to the choice of bin boundaries and readers are cautioned from drawing conclusions based on such representations. We present the observations in this form only to demonstrate visually the generally correct shape of the LF as derived from our model SD. 

\begin{figure}[h] 
   \centering
   \includegraphics[width=6in]{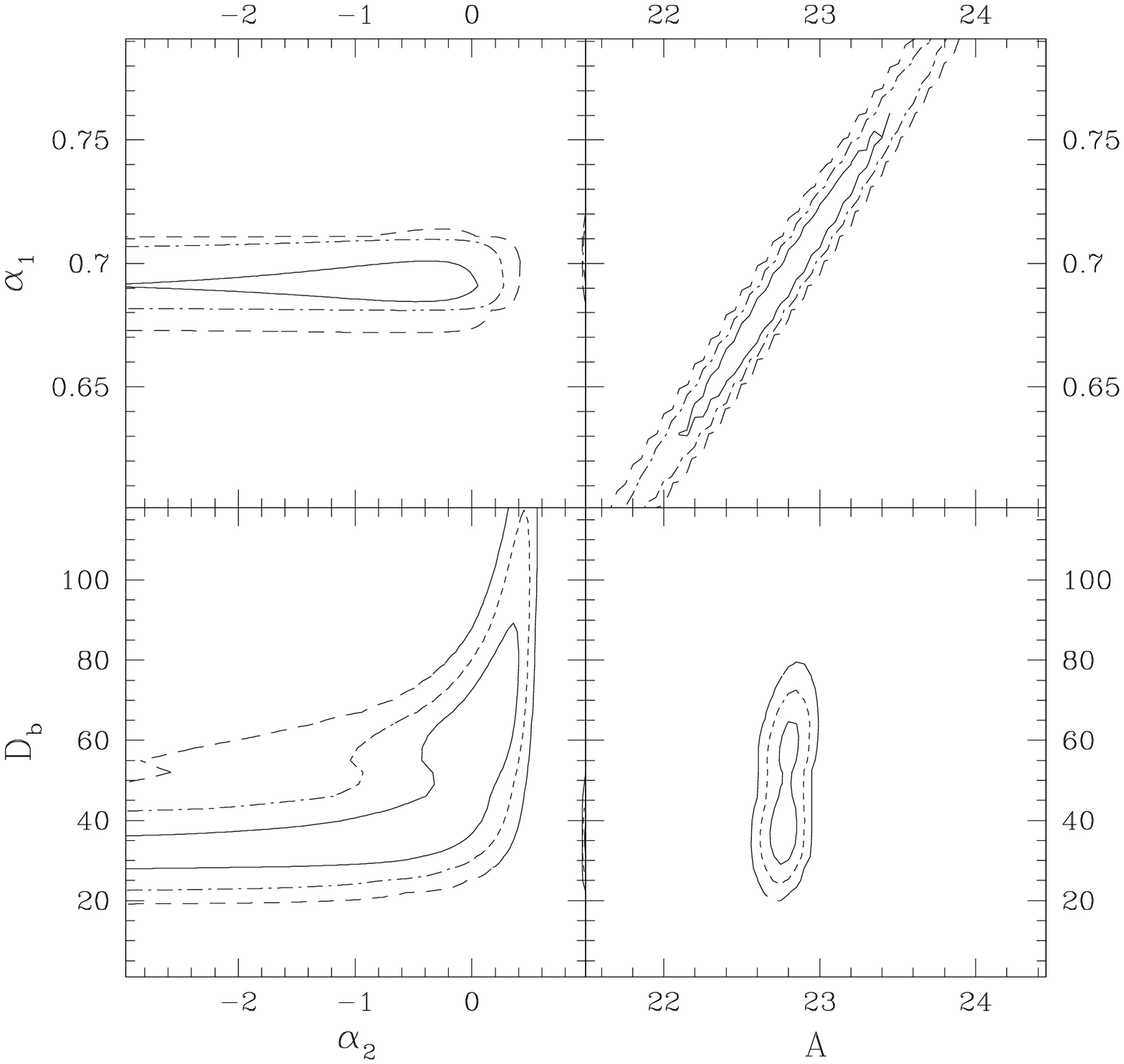} 
   \caption{\label{fig:likelihood}}   
\end{figure}

Fig.~\ref{fig:likelihood}~1, 2, and 3-sigma likelihood contours for pairs of bright and faint object slopes $\alpha_1$, and $\alpha_2$ (or equivalently large and small object size distribution slopes $q_1$ and $q_2$), break diameter $D_b$ (assuming 6\% albedos), and normalization parameter $\log A$. Contours for parameter pairs were generated with the other pair of parameters set to their best-fit values. 

\newpage

\end{document}